\documentclass[10pt]{article}
\usepackage{theorem}
\usepackage{latexsym}
\usepackage[latin1]{inputenc}
\usepackage{amsmath}
\usepackage{amssymb}
\usepackage{amscd}
\clearpage
\usepackage[dvips]{graphicx}
\usepackage{graphicx}

\newcommand{\newc}{\newcommand}

\newc{\be}{\begin{equation}}
\newc{\ee}{\end{equation}}
\newc{\bea}{\begin{eqnarray}}
\newc{\eea}{\end{eqnarray}}
\newc{\beas}{\begin{eqnarray*}}
\newc{\eeas}{\end{eqnarray*}}

\newc{\pardt}{\partial_{t}}
\newc{\pardxi}{\partial_{i}}
\newc{\pardts}{\partial_{t^{*}}}
\newc{\pardxis}{\partial_{i^{*}}}
\newc{\pardxj}{\partial_{j}}
\newc{\pardxk}{\partial_{k}}
\newc{\pard}{\partial}
\newc{\ti}{\tilde}

\newc{\s }{\overline}

\newc{\sect}{\section}
\newc{\subs}{\subsection}

\newc{\defi}{\definition}
\newc{\prop}{\proposition}
\newc{\rem}{\remark}
\newc{\lem}{\lemma}
\newc{\exa}{\example}
\newc{\theo}{\theorem}
\newc{\coro}{\corollary}
\newc{\post}{\postulate}
\newc{\state}{\statement}

\begin{document}
\baselineskip0.5cm
\renewcommand {\theequation}{\thesection.\arabic{equation}}

\title{Effective thermal conductivity  of helium II: from Landau to Gorter-Mellink regimes}

\author{M. Sciacca,$^1$   D. Jou,$^2$ and
M.S.~Mongiovi$^3$}

\date{}
\maketitle
\begin{center} {\footnotesize
$^1$ Dipartimento Scienze Agrarie e Forestali, Universit\`a di Palermo, Viale delle
Scienze, 90128 Palermo, Italy \\
$^2$ Departament de F\'{\i}sica, Universitat Aut\`{o}noma de
Barcelona, 08193 Bellaterra, Catalonia, Spain \\
$^3$ Dipartimento DEIM, Universit\`a di
Palermo, Viale delle Scienze, 90128 Palermo, Italy}
\end{center} \footnotetext{E-mail addresses: michele.sciacca@unipa.it (M. Sciacca),
david.jou@uab.cat (D. Jou), m.stella.mongiovi@unipa.it (M. S.
Mongiov\`{\i}).}

\begin{abstract}
The size-dependent and  flux-dependent effective thermal conductivity
of narrow channels filled with He II is analyzed. The classical Landau evaluation of the effective thermal
conductivity of  quiescent He II 
 is extended  to describe the transition to fully  turbulent regime, where the
 heat flux is proportional to the cubic root of the temperature gradient (Gorter-Mellink regime). To do so we use an  expression for
the quantum vortex line density $L$ in terms of the heat flux considering  the
influence of the walls. From it, and taking into account the
friction force of normal component against the vortices, we compute the effective
thermal conductivity. 

\end{abstract}

\section{Introduction}

Heat transport in small systems or in systems with microscale parts is an active frontier in technology, transport theory,
non equilibrium thermodynamics and statistical mechanics. One of the paradigmatic situations is
the analysis of heat transport along very thin and long wires, whose radius is comparable to the
mean free path of the heat carriers --- as for instance in silicon nanowires with phonons as heat carriers.
In this paper we consider  a narrow  channel filled with He II,
and analyze its transport properties --- namely its effective  thermal conductivity --- in terms of the heat flux and the radius.

The high thermal conductivity of superfluid liquid helium (He II) makes it
an excellent coolant  material, with an important number of
applications as, for instance, the refrigeration of
superconducting magnets in particle accelerators or space cryogenics \cite{Sciver2012, Jones_Cry18, cryogenics}. The interest on the
transport properties of He II in a thin or  very thin channels (for instance diameter from 1 mm to 50 $\mu$m) was in fact an advanced forerunner 
of the later general interest in microfluidics \cite{Bruus}--\cite{GBF_AIP}.
Here, we study the effective thermal conductivity of cylindrical
microchannels filled with He II, a topic   of interest in refrigeration of small systems, in the
behavior of porous systems, and in the research on the effects of
the walls on the quantized vortex lines typical of superfluid
turbulence,  which is a topic of fundamental interest.

In the simplest computation of the effective thermal conductivity of
He II, the resistance to the flow is assumed to be due
to the viscosity of the normal component \cite{Landau}. This leads to a
heat flux which is  proportional to
the temperature gradient (Landau regime). However,
when the heat flux is high enough, quantized vortices appear and  form a vortex tangle  which 
contributes to the overall resistance of the flow \cite{Vinen_PRSA240}--\cite{TsubotaPR}. This implies a
drastic reduction of the effective thermal conductivity, and a strong departure with respect to Fourier's law, since the heat flux becomes proportional
to the cubic root of the temperature gradient (the so-called Gorter-Mellink regime).
In the practice, this increase in thermal resistance may have dramatic consequences if, because of the sudden loss 
of cooling ability, the helium temperature crosses the lambda temperature (about $2.17$ K) and the helium is no longer a superfluid but a normal fluid. Thus, a detailed analysis from Landau regime to Gorter-Mellink regime is relevant on practical grounds. 

Here, instead of  fully developed turbulence, a well-known topic in helium cryogenics,
we  focus our interest on the transition regime from laminar or Landau regime to  turbulent or Gorter-Mellink regime.
This is a topic of practical and theoretical interest, since an exact mathematical description of it is not given often, and the physical understanding is not yet sufficiently developed. For instance, one of the aspects that are not yet well known is the influence of the walls on the quantized vortex tangle of
turbulent superfluid, a topic which is especially relevant in narrow channels.
This is the aim of our paper. 

In particular, we
are interested in analyzing an expression for an effective thermal conductivity $K_{eff}(T,R,L,\dot Q)$  in terms of the radius $R$ of the cylinder,
the temperature $T$, the vortex line density $L$, and the heat current $\dot Q$. The first arguments
on the thermal  conductivity of He II raised in the half of the previous century \cite{Keesom38}--\cite{Mendelsohn}. In 1956 Mendelsohn
reviewed the main experimental peculiarities of He II \cite{Mendelsohn}, and in particular the heat conductivity  in channels
for zero mass flow (he also dealt with nonzero mass flow, but here we are interested  on the former one).  As pointed out by Mendelsohn,  heat conductivity doesn't follow the classical behavior and it seemed to depend on the applied heat flux. It was discovered later that this dependence and the strange behavior of He II is addressed to the
presence of vortex line density. But, at that time the quantized vortex lines were unknown and it was  observed  that the ratio between the gradient of temperature and the heat current,  $\Delta T/\dot Q$, was not longer constant  for an applied heat flux higher than a critical value, and that it was proportional to the second power of the heat current $\dot Q$.  Mendelsohn \cite{Mendelsohn} proposed a theoretical expression for  $\Delta T/\dot Q$ which comes from the formula of the mutual friction force when inserted in the London's formula.  
 which describes  the laminar and the full turbulent regimes, but it is not so accurate nor physically clear in the transition  between both regimes.  

  
The aim of the current paper is  to analyze the relation between $\Delta T$ and $\dot Q$ in terms of the vortex line density $L$ in  the transition between laminar and fully turbulent regimes. The results will be obtained  in terms
of the  one-fluid model (with internal variables) of Extended Thermodynamics \cite{Mongiovi1993} as well as in the two-fluid model \cite{Landau, Donnelly, Barenghi_libro}.
This topic has been considered from the experimental point of view (see for instance \cite{Bartman_Cryogenics, MartinTough} but here we relate it to a theoretical model on the variation between 
$L$ and $\dot Q$ in narrow channels, by means of a generalized Vinen's equation.    

In section 2 we deal with the laminar situation and we compare it with Landau and Tisza \cite{Landau, Tisza_Nature} two-fluid model for the evaluation of the
effective thermal conductivity of He II in a
cylindrical channel and  evaluate such conductivity in
the presence of quantum turbulence, taking into
account the vortex resistance, which for fully developed turbulence corresponds to Gorter-Mellink regime. Section 3 is the original part of this paper: starting from an evolution equation for the vortex line density incorporating the effects of the walls, we propose a mathematical description of the transition regime. Section 4 is devoted to conclusions and remarks.

\section{Effective thermal conductivity of He II in cylindrical channels}
\setcounter{equation}{0}

In this section we deal with the effective  thermal conductivity of He II  along a cylindrical duct. We describe heat transport in terms of the one-fluid model of Extended Thermodynamics \cite{Mongiovi1993} as well as the  Landau-Tisza two-fluid model  \cite{Landau, Tisza_Nature}. 
The basic results of this section are already known, but it is convenient to recall them to make this paper sufficiently self-contained and understandable for a general reader.

If  He II is globally at  rest, the motion of the normal component is
compensated by an opposite flow of the superfluid component, in
such a way that the net velocity of the total system vanishes,
i.e. there is no net mass flow. This requires that  at any time $t$,  $\rho_s \bar
v_s+\rho_n \bar v_n =0$, where $\bar v_s$ and $\bar v_n$ are the
average velocities of the superfluid and normal component on the transversal section of the tube.
This
situation is called {\it counterflow} in literature on He II \cite{Donnelly, TsubotaPR}, and the relevant
quantity here is the so-called counterflow velocity $v_{ns}$, given by
\be%
 v_{ns}=\bar v_{n}-\bar  v_{s}=\frac{\rho}{\rho_s}{\bar v}_{n}.%
\ee

The second equality of the former equation directly follows from
the mentioned condition of vanishing mass flow, namely $\rho_n \bar v_{n}+\rho_s \bar v_{s}=0$. Note for further
use that the heat flow is given by $\bar
q=\rho ST \bar{v}_n=\rho_sST{v}_{ns}$.

Let's assume that the net mass flow is zero (namely the velocity field ${\bf v}$ is small enough  to set ${\bf v} \approx {\bf 0}$).  According to the one-fluid model with the heat flux ${\bf q}$ as internal variable \cite{Mongiovi1993},
the dynamical equations in the stationary situation become

\be
\label{eq:heat_fluxa}
\frac{\pard q_j}{\pard x_j}=0,
\ee
\be
\label{eq:heat_fluxb}
\frac{\pard }{\pard x_j} \left(p \delta_{ij}+m_{<ij>}\right)=0,
\ee
\be
\label{eq:heat_fluxc}
m_{<ij>}= 2\beta T\eta  \frac{\pard q_{<i}}{\pard x_j>},
\ee
\be
\label{eq:heat_fluxd}
  \lambda_1 \frac{\pard T}{\pard x_i} - \beta T^2  \lambda_1  \frac{\pard }{\pard x_j} m_{<ij>}=\sigma_i^q,
\ee
where $\beta$ is a coefficient, ${\bf \sigma}^q$ is the  production term of the heat flux, $p$ is pressure, $m_{<ij>}$ the flux of heat flux, $\eta$ is the shear viscosity, and   $\lambda_1$  can be  interpreted as the heat conductivity   when applied to a  fluid with ${\bf \sigma}^q=- {\bf q}$ \cite{Mongiovi1993}. In these equations, the time derivatives of the corresponding quantities have been neglected because
we are interested in steady-state situations.

\subsection{Laminar situation: Landau regime}

In this subsection  we assume that ${\bf \sigma}^q=- {\bf q}$ while a more general assumption which takes into account of the presence of vortices  will be the argument of the next subsection. 
After some trivial substitutions, and neglecting the nonlinear terms (as for instance terms like $2 \lambda_2 \frac{\pard q_{<i}}{\pard x_j>} \frac{\pard }{\pard x_j}  \left(\beta T\right)$),  equations\;(\ref{eq:heat_fluxb}) and (\ref{eq:heat_fluxd})  become
\be
\label{eq:heat_flux2a}
\nabla  p+\eta \beta T \nabla^2 {\bf q}   =0, \\
\ee
\be
\label{eq:heat_flux2}
\nabla T + \beta T^2  \nabla p=\frac{1}{\lambda_1}{\bf q}.
\ee
Thermal conductivity $\lambda_1$   is linked to the velocity of second sound $w_2$ by the relation $\zeta=\lambda_1/\tau_1= w_2^2 \rho c_V$, where $c_V$ is the constant volume specific heat and $\tau_1$
the relaxation time of the heat flux \cite{Mongiovi1993}. It is experimentally observed that $\lambda_1$ is very high so that the right hand side of the second equation in (\ref{eq:heat_flux2})
may be taken as zero. When (\ref{eq:heat_flux2}) is applied to a cylindrical pipe filled with He II it follows that the 
(\ref{eq:heat_flux2a}) is integrable along the pipe with a constant applied $\nabla p$.
Assuming  that ${\bf q}$  depends only on the radius $r$ one finds
  \be\label{eq:heat_flux3}
 q(r)=-\frac{\Delta p R^2}{4 l \beta T \eta} \left[1-\frac{r^2}{R^2}\right],
 \ee
where $R$ and $l$ are the radius and the length of the pipe, and  the heat flux along the wall has been assumed  to be zero. Taking into account   (\ref{eq:heat_flux2}), with vanishing right-hand side, 
namely  
$ \Delta p=-( \beta T^2)^{-1} \Delta T$, the mean value of the heat flux over the transversal section of the pipe obtained from (\ref{eq:heat_flux3})  may be expressed  in terms of $\Delta T$ as
\be\label{eq:heat_flux_mean2}
\bar q=-\frac{\Delta p R^2}{8 l \beta T \eta}=\frac{\Delta T R^2}{8 l \beta^2 T^3 \eta}.
\ee

Therefore, since the total heat flux across   the transversal area is $\dot Q=\pi R^2 \bar q$, 
 the effective heat conductivity according to the Fourier's law is
\be\label{eq:conductivity1}
K_{eff}=\frac{\dot Q}{\pi R^2}\frac{l}{\Delta T} =\frac{R^2}{8  \beta^2 T^3 \eta}=\frac{R^2 \rho^2 T S^2}{8 \eta}.
\ee
We have used $\beta=-(\rho S T^2)^{-1}$ \cite{Mongiovi1993} to make evident that  (\ref{eq:conductivity1})  is  the well-known Landau formula for thermal conductivity \cite{Sciver2012, Landau, Mongiovi2001, Jou-book}.
Note that  (\ref{eq:conductivity1})  is not  a true thermal conductivity, dependent only on the material, but a global quantity, that depends quadratically on
the radius $R$ of the cylinder, and therefore it is
 strongly reduced for thin capillaries.
 
\subsection{Turbulent situation: Gorter-Mellink regime}

Let's now take into account  the presence of quantum vortices in He II, when the heat flux becomes higher than a threshold value. The presence of the vortices
is described by the vortex length density $L$, and their main effect on the flow is an internal friction
between the vortices and the normal fluid \cite{Vinen_PRSA240}--\cite{TsubotaPR}.

In the one-fluid model \cite{Mongiovi1993, Jou-book, MJ-2005a}, the effects of the internal friction are described through a contribution to the production term ${{\bf \sigma}^q}$ appearing
in equation\,(\ref{eq:heat_fluxd}), which is taken now as
  ${\bf \sigma}^q=-{\bf q}-\tau_1 K L {\bf q}$, where $K=\frac{1}{3}\kappa B_{HV}$, $B_{HV}$ being the dimensionless Hall-Vinen friction coefficient, $\kappa$ the quantum  of vorticity $h/m$ (with $m$ the mass  of helium atom and $h$ the Planck's constant 
in such a way that $\kappa=9.97\times 10^{-8}$m$^2/$s),   and $\tau_1$ the relaxation time of the heat flux. Hence
 equations\;(\ref{eq:heat_flux2a}) and (\ref{eq:heat_flux2}) in the stationary case become
\bea
&&\label{eq:heat_flux4} \nabla p+
\eta \beta T \nabla^2 {\bf q}   =0,\\
&& \label{eq:heat_flux4b}  \nabla T + \beta T^2  \nabla p=-\frac{1}{\lambda_1}{\bf q}- \frac{K L}{\zeta} {\bf q},
\eea
where $\zeta=\lambda_1/\tau_1$ as said below equation\,(\ref{eq:heat_flux2}).
Assuming that  $\lambda_1$ is very high to neglect the first term in the right hand side of (\ref{eq:heat_flux4b}), but comparable to the relaxation time $\tau_1$ in such a way that $\zeta$ is finite 
(it also determines the second sound velocity), 
equations\;(\ref{eq:heat_flux4}) and (\ref{eq:heat_flux4b}) reduce to
\bea 
&&\label{eq:heat_flux5a} \nabla p+\eta \beta T \nabla^2 {\bf q}   =0,\\
&&  \label{eq:heat_flux5b} \nabla T + \beta T^2  \nabla p+ \frac{K L}{\zeta} {\bf q}=0.
\eea

The solution of (\ref{eq:heat_flux5a}) applied to He II in a cylindrical pipe  is still  (\ref{eq:heat_flux3}) and the mean value
is (\ref{eq:heat_flux_mean2}). By integrating the second equation of (\ref{eq:heat_flux5b}) along the pipe  we find
\be\label{delta}
\Delta T + \beta T^2  \Delta p=\frac{K l \bar L}{\zeta} \bar q,
\ee
since $\nabla T=-\Delta T/l$,  $\nabla p=-\Delta p/l$
and using (\ref{eq:heat_flux5a}), in view of the previous identification of $\beta$ as $\beta=-(\rho s T^2)^{-1}$, and writing $\Delta p$ in terms of $\bar q$ one obtains
 \be\label{delta2}
 \Delta T=\frac{8\eta l}{R^2 \rho^2 S^2 T}\bar q+\frac{K l \bar L}{\zeta} \bar q.
 \ee
 From here, it follows that the effective thermal conductivity is
\be\label{delta3}
K_{eff}=\frac{R^2 \zeta}{8\beta^2T^3\eta\zeta+K \bar L R^2}.
\ee
 Equation (\ref{delta2}) may also be written in terms of $\dot Q$ as
  \be\label{delta2bis}
 \Delta T=\frac{8\eta l}{\pi R^4 \rho^2 S^2 T}\dot Q+\frac{Kl \bar L}{\pi R^2 \zeta} \dot Q.
 \ee
The first term corresponds to Landau regime; if one takes $L\sim \dot Q^2$, the second term corresponds to the Gorter-Mellink regime. We want to study the transition from one to the other.
 
It must be emphasized that Landau's expression  is valid for fully developed Poiseuille flow, i.e. for sufficiently long channels in which the entrance region is negligible as compared with whole length. For short enough
channels, the fluid has no time enough to reach the classical parabolic velocity profile,
and the equations must incorporate nonlinear terms corresponding to the convective derivative. In particular,
Lesniewski et al. \cite{Les1996} investigated on the influence of the inertia effects
in short narrow ducts and they proposed the Shah-London equation for the boundary layer assessment in laminar flows.
According to them, the Poiseuille flow is established only for sufficiently long tubes, but
a precise estimation of the critical length is not given because  the experimental data are very scarce.
They  take into account  two experiments by Childers and Tough, for which they show
a $1/3$ factor and $2/3$ factor in log-log scale in $\dot Q$ vs $\nabla T$ (the first corresponding to quantum turbulence according to
the Mendelsohn's proposal).
From the figure plotted by the authors, the dependence of the critical length for the
establishing Poiseuille flow to the applied heat flux is grasped.
 
 \section{Transition from Landau to Gorter-Mellink regime}
\setcounter{equation}{0}

Our aim is to propose a mathematical description for the transition from Landau to Gorter-Mellink regimes. Thus   instead of directly taking $L \propto \dot Q^2$ in (\ref{delta2bis}), which is typical of fully developed turbulence in wide channels, we pay  attention to a more detailed relation between $L$ and $\dot Q$ in narrow channels.
 
In fact, the mentioned transition implies a narrow intermediate regime (TI turbulence)  \cite{Donnelly}--\cite{TsubotaPR}, with a
relatively low vortex line density and a transition to a more developed turbulence ( TII turbulence).
According to experimental results
\cite{MartinTough}  a relevant quantity in such transition is the quantum Reynolds number $v_{ns} d/\kappa$,   with $v_{ns}$ the counterflow velocity, 
$d$ the diameter of the channel, and  $\kappa$ the quantum vorticity $h/m$ ($h$ the Planck constant and $m$ the helium mass).
Since $\kappa$ has dimensions  of (length)$^2$/time it plays in $v_{ns}d/\kappa$ a role analogous
to that of kinematical viscosity $\nu$ in classical Reynolds number $V d/\nu$, with $V$ the velocity \cite{Jou_BMPA5}.

For instance,  the transition from laminar regime   to the first kind of turbulence (TI turbulence)
 is at $Rey_1=127$ at $T=1.5$ K (for
$\dot Q=5\times 10^{-4}$ J/s the diameter for which this transition will be
detected is $d=1330\; \mu$m, and for $\dot Q=10\times 10^{-4}$ J/s, $d=2650\;
\mu$m)  \cite{Jou_BMPA5}. A further transition to TII turbulence   is
found  at quantum Reynolds number $Rey_2=226$, which yields the values $d=740\; \mu$m
(for $\dot Q=5\times 10^{-4}$ J/s) and $d=1490\; \mu$m (for $\dot Q=10\times 10^{-4}$ J/s).
At $T=1.6$ K the respective values of the mentioned critical  quantum Reynolds numbers are $Rey_1=112$ and $Rey_2=212$, and  at $T=1.7$ K
they are $Rey_1=96$ and $Rey_2=187$. Note the dependence of these values on the temperature. All these complexities must be taken into account in order to describe the transition from Landau to Gorter-mellinck regime, as we do in this section.
 
  \subsection{Relation between vortex length density and heat flux $q$ in narrow channels}

 The classical Vinen's equation for the evolution of $L$ is \cite{Vinen_PRSA240}--\cite{TsubotaPR}
 \be\label{Vinen}
 \frac{d L}{d t} = \alpha_v  v_{ns} L^{3/2}-\beta_v \kappa L^2,
 \ee
 with $\alpha_v$ and $\beta_v$ dimensionless coefficients which depend on $T$ and which are related to vortex formation 
 and destruction, respectively. The steady-state result of (\ref{Vinen})  is
 \be
 L^{1/2}=\frac{\alpha_v}{\beta_v \kappa} v_{ns}=\frac{\alpha_v}{\beta_v \kappa \rho_s T S} \bar q,
 \ee
 leading to $L \propto q^2$. This corresponds to fully developed turbulence.

Equation (\ref{Vinen}) does not describe the transition from laminar flow to fully developed
flow. Such a transition may be described by means of a generalized Vinen's equation including the wall effects, which, in the
simplest version is  \cite{MJ-2005a}
  \be\label{Vinen-gen2}
 \frac{d L}{d t} =-\beta_v \kappa L^2+\left[\alpha_0 v_{ns}-\omega^\prime \beta \frac{\kappa}{d}\right] L^{3/2},
  \ee
with $d$ the diameter of the tube, and the coefficients $\alpha_0$ and $\omega$ are functions of $v_{ns} d/\kappa$, the quantum Reynolds
number. The third term takes into account the effect of having a thin tube with diameter $d$ finite and, therefore, it modelizes the influence of the wall. For narrow channels this term --- absent
from (\ref{Vinen}) --- becomes especially relevant, whereas it becomes negligible for wide channels.

Equation (\ref{Vinen-gen2}) has  the steady state solutions
\be\label{Lsolution}
L=0;   \qquad L^{1/2}= \frac{\alpha_0}{\beta \kappa} v_{ns}-\frac{\omega^\prime}{d}.
\ee
The non zero solution  exists and is stable for $v_{ns}>V_{c1}=\frac{\beta \kappa \omega^\prime}{\alpha_0 d}$ . From the experimental results reported in \cite{MartinTough}  it is seen  that the second kind of solution has
two different regimes, namely a TI turbulence for  a quantum Reynolds number $Rey=v_{ns} d/\kappa$, between $Rey_1$ and $Rey_2$ , described by  \cite{MartinTough} 
\be\label{TI}
  L^{1/2} =\frac{\gamma_{TI}}{\kappa} v_{ns}- 1.48 \frac{\alpha_1}{d},
\ee
and TII turbulence flow for $Rey>Rey_2$ described by  \cite{MartinTough} 
\be\label{TII}
  L^{1/2} =\frac{ \gamma_{TII} }{\kappa} v_{ns}-1.48 \frac{\alpha_2}{d},
\ee
 with $\gamma_{TI}$, $\gamma_{TII}$ and $\alpha_i$ numerical constants which depend on temperature and whose values are reported in Table 1 \cite{MartinTough}.
The second solution in  (\ref{Lsolution}) fits the experimental data \cite{MartinTough} in the TI regime for $\frac{\alpha_0}{\beta}=\gamma_{TI}$ and $\omega^\prime=1.48 \alpha_1$, whereas in the TII regime for $\frac{\alpha_0}{\beta }=\gamma_{TII}$ and $\omega^\prime=1.48 \alpha_2$.

The  transition from the TI turbulent regime to the TII turbulent regime can be described in (\ref{Vinen-gen2}) by
assuming that coefficient
$\alpha_0$ depends on the quantum Reynolds number $Rey$ as \cite{MJ-2005a}
\be\label{gamma0}
\gamma_0(Rey)=\frac{\alpha_0(Rey)}{\beta }=\alpha_{c}\left(1+c \tanh \left[A\left(Rey-Rey_2\right)\right]\right)
\ee
in such a way that $\gamma_0(Rey)= \gamma_{TI}$ for $Rey_1 \ll Rey \ll Rey_2$   and   $\gamma_0(Rey)= \gamma_{TII}$ for $Rey \gg Rey_2$,
with  $\alpha_{c}=\frac{\gamma_{TI}+\gamma_{TII}}{2}$ and $c=\frac{\gamma_{TII}-\gamma_{TI}}{\gamma_{TI}+\gamma_{TII}}$. The coefficient $A$ is
chosen  to fit better the transition regime from turbulence TI and turbulence TII. In Figures below we have chose the expression $1.47/(Rey_{edge}-Rey_2)$, where $Rey_{edge}=V_{edge} d/\kappa$ with $V_{edge} $ being the counterflow velocity  which guarantees the 90\% of the codomain of $\tanh$  between the edges of the transition interval.

Also, a  similar expression is required for the coefficient $\omega^\prime$ \cite{MJ-2005a}
\be\label{omega}
\omega^\prime(Rey)=\beta_{c}\left(1+c_1 \tanh \left[A\left(Rey-Rey_2\right)\right]\right)
\ee
with $\beta_c=0.74 (\alpha_1+\alpha_2)$ and $c_1=\frac{\alpha_2-\alpha_1}{\alpha_1+\alpha_2}$, in such a way $\omega^\prime=1.48 \alpha_1$ in the TI regime and $\omega^\prime=1.48 \alpha_2$ in the TII regime.

\begin{figure}[h]
\begin{center}
\resizebox{4.93cm}{!}{\includegraphics{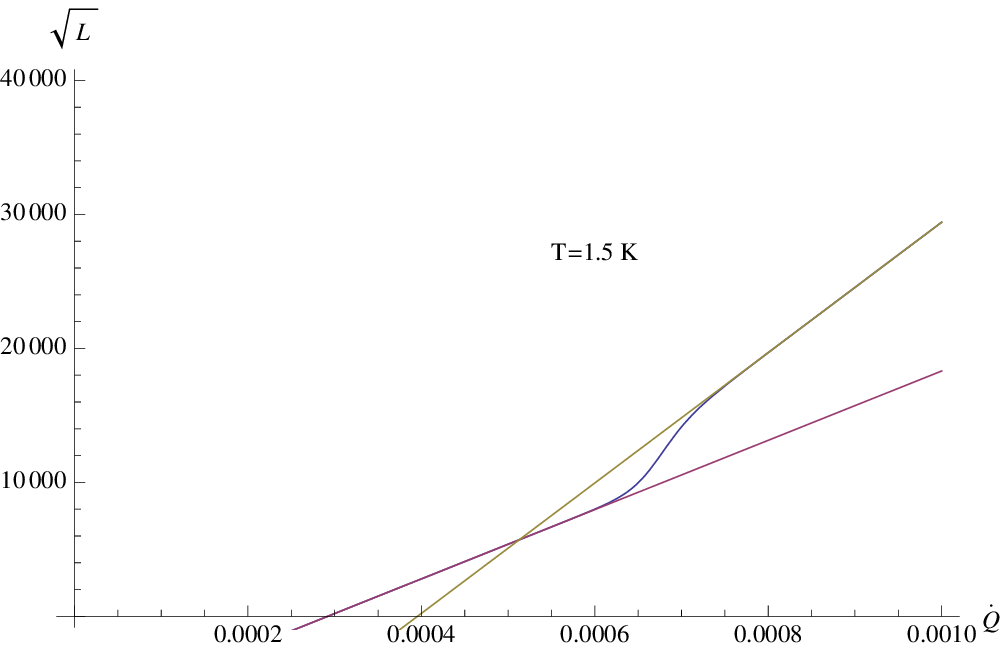}} \quad
\resizebox{4.93cm}{!}{\includegraphics{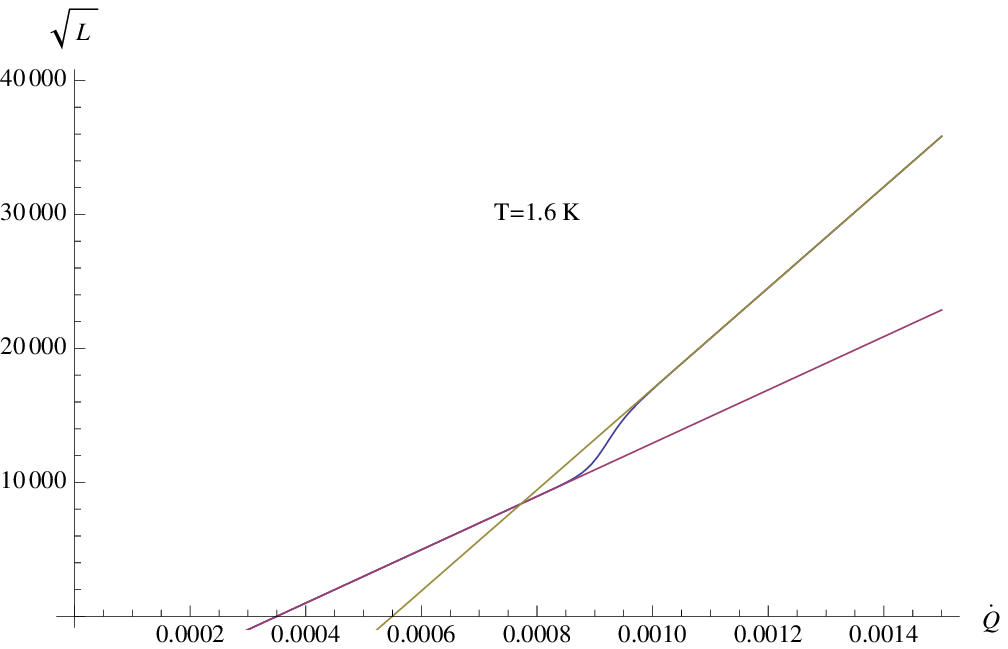}}\quad 
\resizebox{4.93cm}{!}{\includegraphics{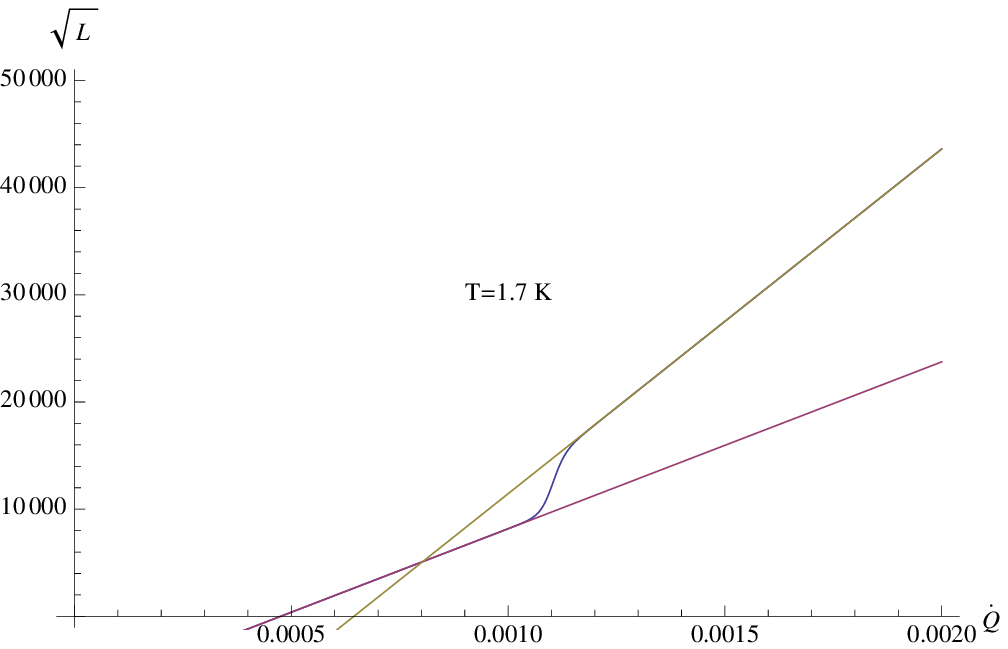}}
\end{center}
 \caption{[Color online]   Plot of $L^{1/2}$ (m$^{-1}$) vs the heat flux $\dot Q$ (J/s) for three temperatures $T=1.5$\,K (left), $T=1.6$ K (middle) and $T=1.7$ K (right). In each figure the two regimes TI (yellow) and  TII (bordeaux), given by the straight lines are compared with our solution (\ref{Lsolution}b) with $\gamma_0$
 and $\omega^\prime$ given in (\ref{gamma0}) and (\ref{omega}) and for $d=10^{-3}$ m.}\label{vortexline}
\end{figure}

In Figures\;\ref{vortexline} the  second solution  (\ref{Lsolution}) with (\ref{gamma0}) and (\ref{omega}) (the blue line) is compared to the experimental results  by Martin and Tough \cite{MartinTough} for the TI regime (\ref{TI}) (the yellow line) and for the TII regime (\ref{TII}) (the bordeaux line). Plots refer to three different temperatures: $T=1.5$ K (the left figure), $T=1.6$ K (the middle figure), and $T=1.7$ K (the right figure).
 
 In the next section we directly go to the experimental results obtained by Tough and collaborators on heat transfer in He II in cylindrical channels\cite{MartinTough}. For high $v_{ns}$, equation\;(\ref{TII}) may be approximated by $L^{1/2} = \gamma_{TII} v_{ns}$, and (\ref{delta2bis})
 becomes analogous to  the Mendelsohn's proposal, but for intermediate values of $v_{ns}$ corresponding to $Rey_1<Rey<Rey_2$ and slightly above $Rey_2$, equations\;(\ref{TI}) and \;(\ref{TII}) must be used.

\subsection{Quantum turbulence: explicit evaluation}

In the previous sub-section we have seen that thermal conductivity
depends on the vortex line density, which in wide channels is
related to the counterflow velocity in a very direct and simple
way. However, in narrow channels, $L$ depends also on the diameter
of the channel, as shown by Tough and collaborators in their
studies over the 1980's \cite{MartinTough}. This happens when
radius becomes comparable to the average separation of vortex
lines, which is of the order of $L^{-1/2}$. They obtained that
He II is laminar without vortex
lines, for $Rey< Rey_1$ (except  some remaining vortex lines of previous
experiments, pinned to the walls or formed at the
$\lambda$-transition); for $Rey_1< Rey< Rey_2$ there
is the so called turbulence $TI$: a mild form of turbulence
characterized by a relatively low value of $L$ (\ref{TI}); for
$Rey>Rey_2$ there is a steep increase in $L$, and
the value of $L$   (\ref{TII})  increases for increasing $Rey$.
 The values of the critical Reynolds numbers depend on the temperature and they   are
reported in Table\,\ref{TableRey}. Here we assume that the values of these Reynolds numbers for the transition only depend on temperature. This allows to extrapolate the experimental observations in \cite{MartinTough} (made at radii of the order of 1 mm to several mm) to other value of the radii, up to micro metric order, provided the radius is sufficiently bigger than the phonon free path.

\begin{table}[h]
  \centering
\begin{tabular}{|c|c|c|c|c|c|c|c|c|c|c|c|}
  \hline
  $T$ & $\rho$ & $\eta$  & $\rho_s$  & $\gamma_{TI}$ & $B_{HV}$ & $\alpha_1$
  & $\gamma_{TII}$ & $\alpha_2$ & $S$ & $\zeta$ & $A$ \\
  \hline
  1.5 & 145 & $1.35 \cdot
  10^{-6}$ & 129 & $7.68 \cdot
  10^{-2}$  & 1.296 & 5.1 & $14.45  \cdot
  10^{-2}$ & 13 & $196$  & $6.57 \cdot
  10^{3}$ & 0.086\\
  \hline
  1.6 & 145 & $1.3 \cdot
  10^{-6}$  & 122 & $8.57 \cdot
  10^{-2}$ & 1.193 & 4.7 & $16.25 \cdot
  10^{-2}$ & 14 & $282$  & $9.29 \cdot
  10^{3}$ & 0.122\\
  \hline
  1.7 & 145 & $1.29 \cdot
  10^{-6}$ & 112 & $9.17 \cdot
  10^{-2}$ & 1.1 & 5 & $18.94 \cdot
  10^{-2}$ & 14 & $395$  & $13.16 \cdot
  10^{3}$ & 0.21\\
  \hline
\end{tabular}
  \caption{In the table temperature $T$ (K), density of helium $\rho$
  (Kg m$^{-3}$), dynamic viscosity $\eta$ (Kg (s m)$^{-1}$),
  density of the superfluid component $\rho_s$ (Kg m$^{-3})$, specific entropy $S$ (J (Kg K)$^{-1})$, the coefficient  $\zeta$ (J /(m K s$^{2}))$, and dimensionless parameters  $B_{HV}$,  $\gamma_{TI}$, $\alpha_1$, $\gamma_{TII}$ and
  $\alpha_2$ refer to Martin and Tough's experiments
  \cite{MartinTough}, using Ref.\,\cite{Tsubota2010}
  and \cite{Barenghi1998}. }\label{Table3}
\end{table}

\begin{table}[h]
  \centering
\begin{tabular}{|c|c|c|c|}
  \hline
  $T$ & 1.5 K & 1.6 K  & 1.7 K  \\
  \hline
  $Rey_1$ & 127 & 112 & 96 \\
  \hline
   $Rey_2$ & 226 & 212  & 187\\
  \hline
\end{tabular}
  \caption{In the table  the critical quantum Reynolds numbers  $Rey_1$ for the appearance of  TI turbulent regime
  and $Rey_2$ for the appearance of TII turbulent regime for three temperatures are reported from the   Martin and Tough's experiments
  \cite{MartinTough}. }\label{TableRey}
\end{table}

The explanation of this  steep increase of $L$  is still
open to debate because there isn't a definitive proof, even
experimentally, of what these two states, $TI$ and $TII$, are. But,
it is worth mentioning  two different possible explanations: the
one proposed by two of us in Ref.\,\cite{MJ-2005a} and the one
proposed by Melotte and Barenghi in Ref.\,\cite{Melotte}. The
former interpreted the steep increase of $L$ at $Rey_2$ as the
beginning of vortex reconnection, namely, to the production of a
high number of free vortex loops as a consequence of the crossing
and cutting and recombining of vortex lines that in turbulence
$TI$ were most of them pinned to the walls. The increase of $L$ in
turbulence $TI$, instead, is basically due to Kelvin wave
excitations in pinned vortex lines.
The second proposal, instead, of Melotte and Barenghi \cite{Melotte} explains
this transition to the $TII$ state as a consequence of the
transition from laminar to turbulence flow for the normal
component. However, the critical velocity for superfluid turbulence
is of one order smaller than the critical velocity for the normal
component. Of course each proposal does not preclude the other
one, and a combined interplay of the two proposals  is also
possible.

The description of $\Delta T$  in terms of $\dot Q$ will be

\begin{eqnarray}
\Delta T&=&\frac{8 \eta l}{\pi R^4 \rho^2 S^2 T}\dot Q  \qquad \textrm{for} \qquad Rey< Rey_1; \label{L0}\\
  \Delta T&=&\frac{8 \eta l}{\pi R^4 \rho^2 S^2 T}\dot Q +\frac{Kl }{ \zeta}\left[ \frac{\gamma_0}{\kappa \rho_s T S}  \frac{\dot Q}{\pi R^2}- \frac{\omega^\prime}{2 R}\right]^2 \frac{\dot Q}{\pi R^2},  \hspace{0.16cm} \textrm{for} \hspace{0.16cm} Rey>Rey_1; \label{L1}
\end{eqnarray}
where
\be
\gamma_0=\frac{\gamma_{TI}+\gamma_{TII}}{2}\left(1+\frac{\gamma_{TII}-\gamma_{TI}}{\gamma_{TII}+\gamma_{TI}} \tanh \left[A\left(Rey-Rey_2\right)\right]\right)
\ee
\be
\omega^\prime =0.74 (\alpha_1+\alpha_2)\left(1+\frac{\alpha_2-\alpha_1}{\alpha_1+\alpha_2} \tanh \left[A\left(Rey-Rey_2\right)\right]\right)
\ee
where  $Rey$ can be expressed in terms of  $\dot Q$ by $Rey=\frac{2 v_{ns }R}{\kappa}=\frac{2 \dot Q}{\kappa \rho_s T S \pi R}$
and $A$ can be evaluated assuming that the $90\%$  of the codomain of $\tanh$ is between the edges of the transition interval.
 From the second term in the rhs of equation (\ref{L1}) one obtains the critical value
$\dot Q_{c}=\frac{\omega^\prime \kappa \rho_s T S \pi R}{2\gamma_{0}}$.

 These expressions for the vortex line density in narrow channels
allow to obtain the effective thermal conductivity, which now depends not only on the radius but also on the applied heat current $\dot Q$
\begin{equation}\label{8narrow}
    K_{eff-turb}= \frac{T\rho^2 S^2  R^2 \zeta}{8 \eta \zeta+ K T\rho^2 S^2  R^2 \left( \frac{\gamma_0}{\kappa \rho_s T S}  \frac{\dot Q}{\pi R^2}-\frac{\omega^\prime}{2R}\right)^2},
\end{equation}
where the relation $\dot Q=\pi R^2 q=\pi R^2 \rho_s T S v_{ns}$ has been used,
and consequently the ratio between $K_{eff-Landau}$ and
$K_{eff-turb}$ becomes
\be\label{conductivityNarr}%
\frac{K_{eff-Landau}}{K_{eff-turb}}=1+ \frac{K R^2 \rho^2S^2 T}{8 \eta \zeta} \left( \frac{\gamma_0}{\kappa \rho_s T S}  \frac{\dot Q}{\pi R^2}-\frac{\omega^\prime}{2R}\right)^2.%
\ee

\begin{center}
\begin{figure}
  \resizebox{6cm}{!}{\includegraphics{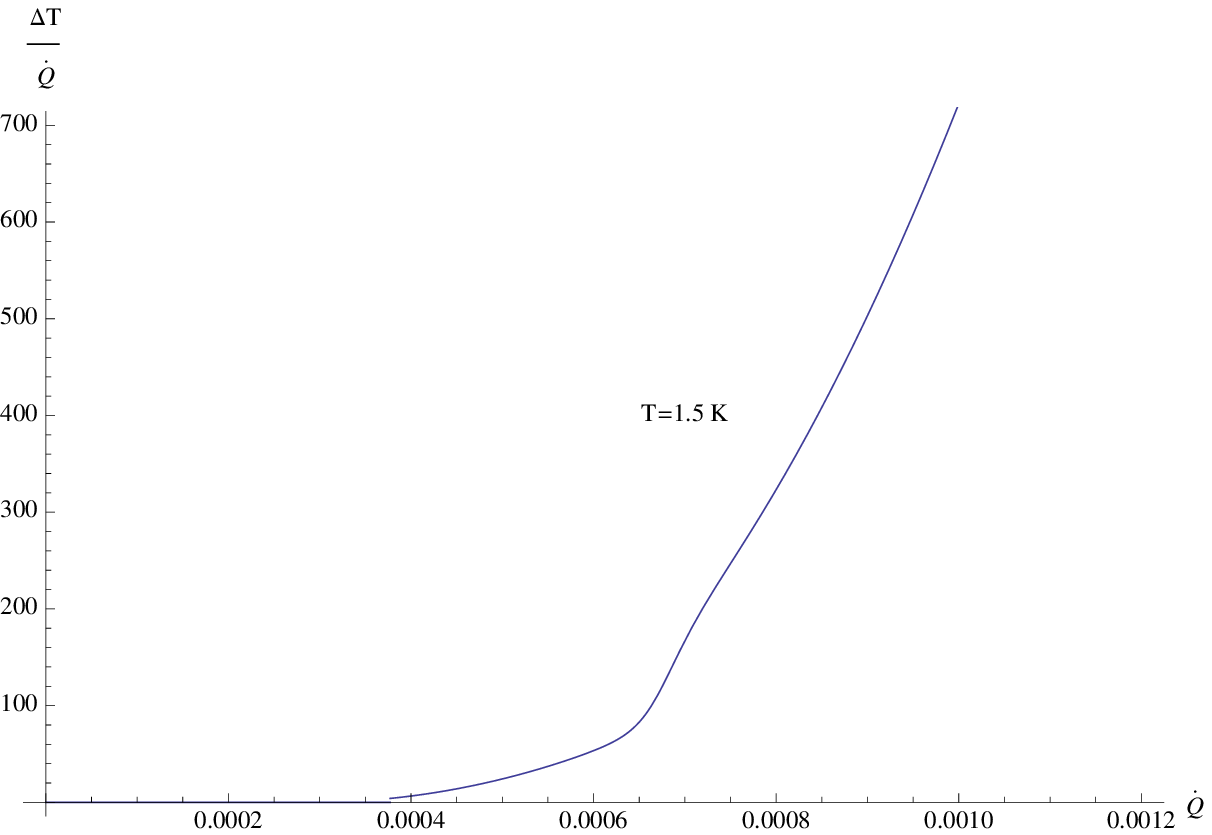}} \qquad
\resizebox{6cm}{!}{\includegraphics{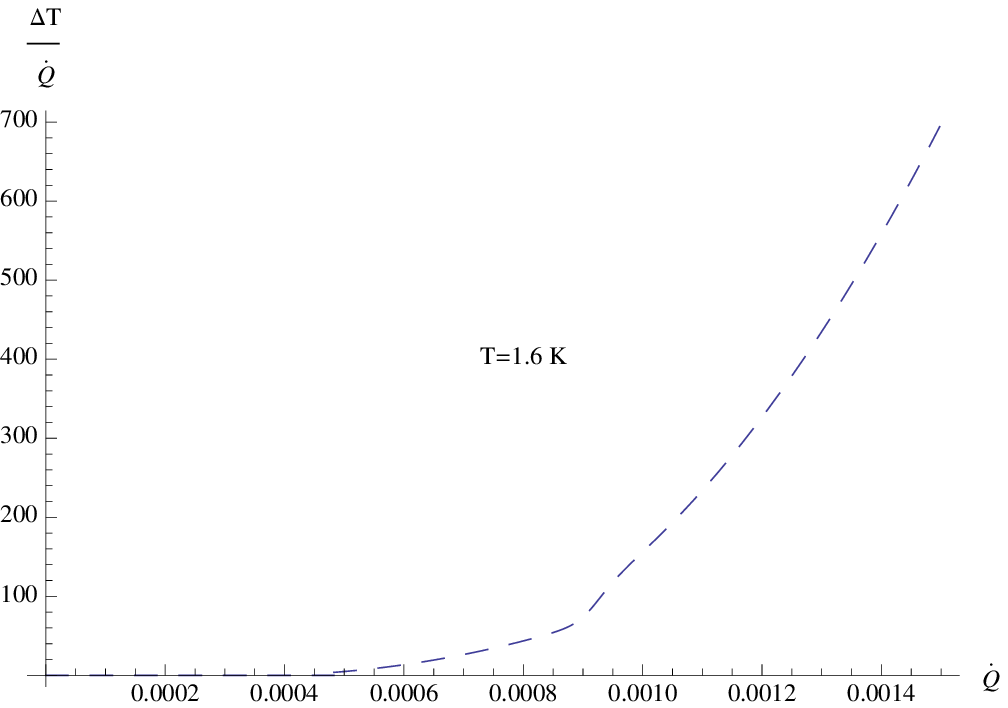}}
\resizebox{6cm}{!}{\includegraphics{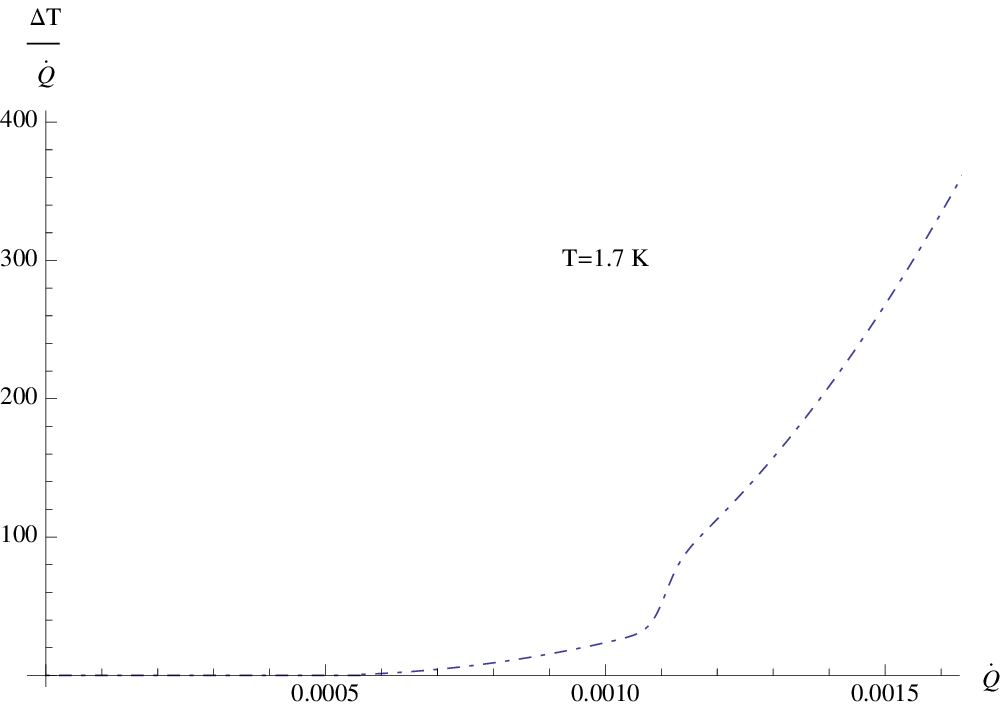}}\hspace{3cm}
\resizebox{6cm}{!}{\includegraphics{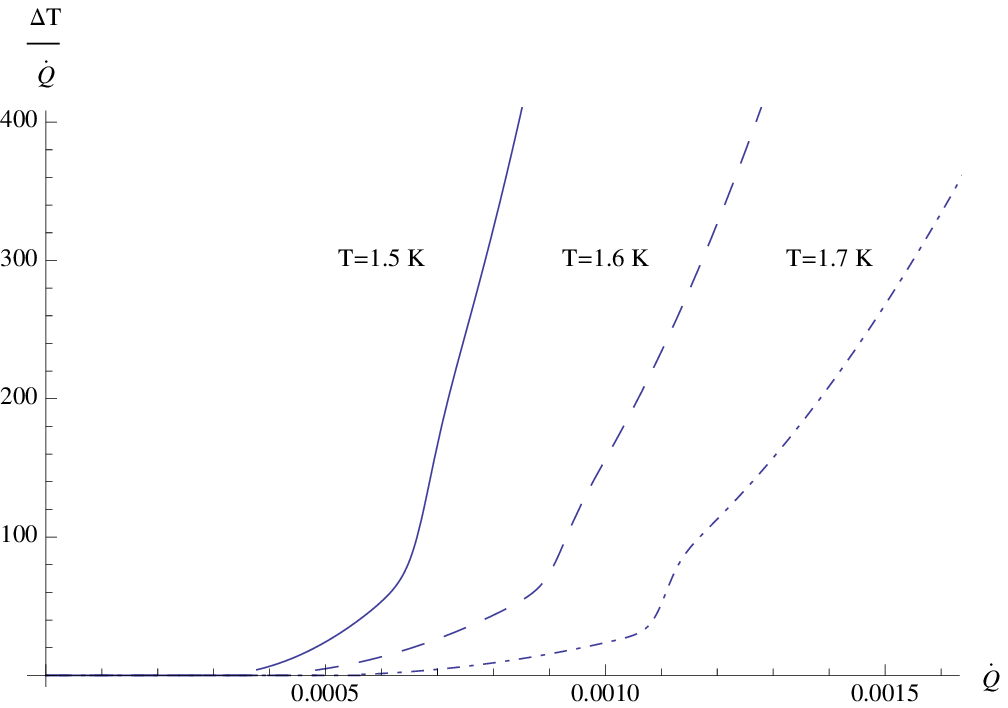}}
 \caption{Behaviour of the ratio $\Delta T/\dot Q$ (K s/J) versus $\dot Q$  (J/s) in superfluid
  helium without vortices (the flat left part of each line) and with vortices in the TI  and TII regimes (the right part). Data are from the Martin and Tough's experiment for radius  $1000\;\mu$m at three different temperatures:  $T=1.5$ K (upper left), $T=1.6$ K (upper right) and $T=1.7$ K (below left), and all temperatures together (below right).}\label{narrow00}
  \end{figure}
\end{center}

We apply our results (\ref{L0})--(\ref{L1}) and (\ref{conductivityNarr}) to the experiments of
Martin and Tough \cite{MartinTough}, who made  systematic
measurements on heat transfer in He II in cylindrical channels. We
take $d=1$ mm, $l=10$ cm and $T=1.5$ K, $T=1.6$ K and  $T=1.7$ K.
In Table \ref{Table3} the values of the parameters used in the
calculations are reported. Thermal conductivities (\ref{conductivityNarr})  from the Martin
and Tough's experiments are drawn in Figure \ref{narrow00} for
three different temperatures ($T=1.5$ K, $T=1.6$ K, $T=1.7$ K)
against the heat current $\dot Q$. For the same values, the ratio $\Delta T/\dot Q$ vs $\dot Q$
is plotted in Figures\;\ref{narrow00}  for (\ref{L0})--(\ref{L1}) in order to compare it to the experimental data
from \cite{Mendelsohn}.

\begin{figure}
  \includegraphics[width=10cm]{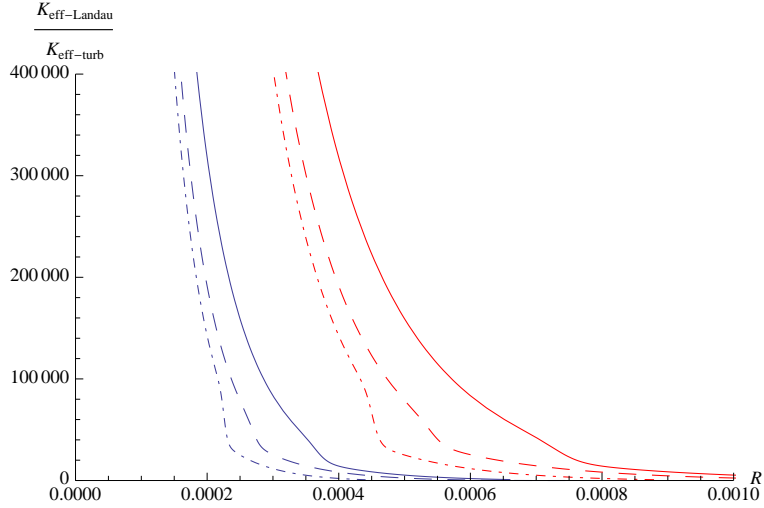}\\  
  \caption{Behaviour of the ratio between thermal conductivity in He II without and with vortices  (\ref{conductivityNarr})  against the radius of the channel (m). The blue line refers to an applied heat current $\dot Q=5\times 10^{-4}$ J/s whereas the red lines refers to an applied heat current $\dot Q=10^{-3}$ J/s. Each case  is considered for  three temperatures:  $T=1.5$ K (solid line), $T=1.6$ K (dashed line) and $T=1.7$ K (dot-dashed line).}\label{narrow2}
\end{figure}

The same expression (\ref{conductivityNarr}) is then plotted in
Figure \ref{narrow2} at two fixed
applied heat flux:  $\dot Q=5\times 10^{-4}$\,J/s and $\dot Q=10\times 10^{-4}$\,J/s for $Rey > Rey_1$.
By writing the quantum Reynolds number in terms of the heat current $\dot Q$, namely $Rey=\frac{2 \dot Q}{\kappa \rho_s T S \pi R}$,
then the turbulent status for an applied heat current in terms of the radius of the channel can be established. Indeed, from
$\frac{2 \dot Q}{\kappa \rho_s T S \pi R}>Rey_1$ we find  $R<\frac{2 \dot Q}{\kappa \rho_s T S \pi Rey_1}$, namely the narrower is the tube the higher is
the turbulent vortex line density $L$. Figure\;\ref{narrow2} shows that in channels with small diameter the effective thermal conductivity is small and decreases with the diameter.
Note that in figure\;\ref{narrow2} a cutoff in the radius of the channel has to be considered because for diameter small enough the quasiparticle of superfluid helium  cannot flow 
through the narrow channel and a ballistic regime is reached, which will not be discussed here \cite{Bartman_Cryogenics}. 

\begin{figure}
  \includegraphics[width=10cm]{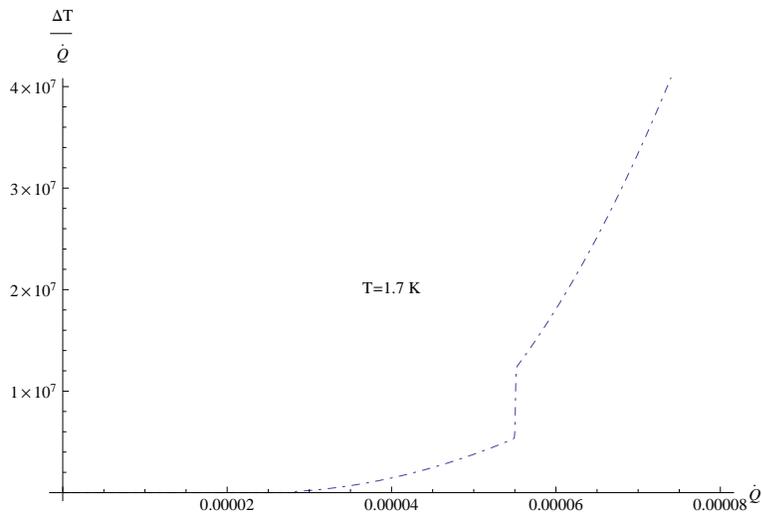}\\  
  \caption{Behaviour of the ratio $\Delta T/\dot Q$ (K s/J) versus $\dot Q$  (J/s) in He II for a microchannel filled with helium II  with diameter 50 $\mu$m at $T=1.7$ K.}\label{narrow3}
\end{figure}

\begin{figure}
  \includegraphics[width=10cm]{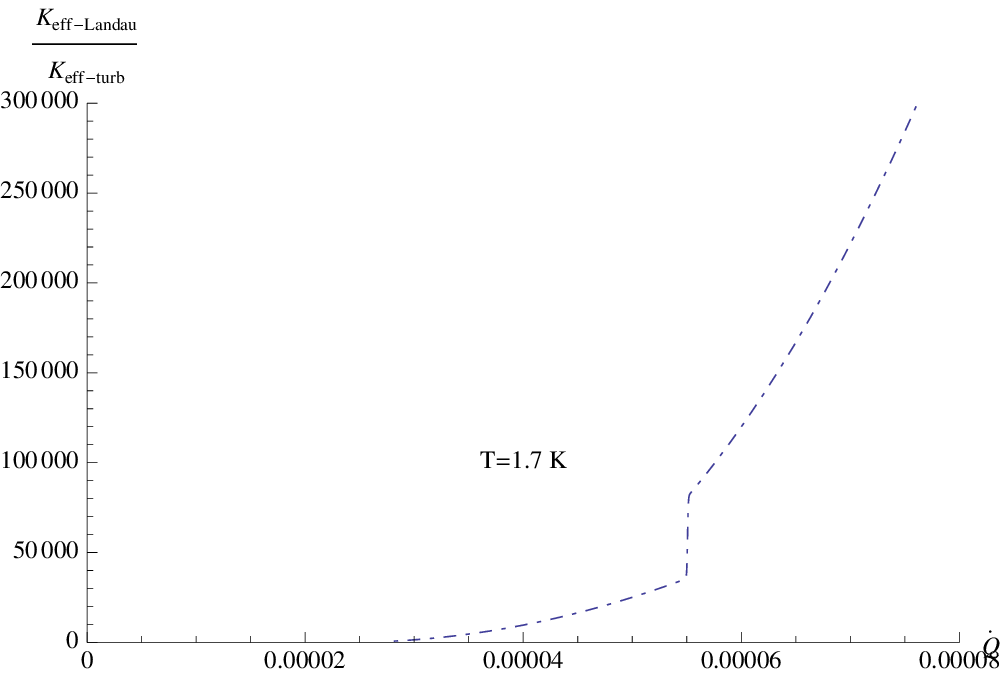}\\  
  \caption{The behaviour of the ratio between thermal conductivity in He II without and with vortices  (\ref{conductivityNarr}) against the applied heat current $\dot Q$  (J/s) in a micro channel with a diameter of 50 $\mu$m at $T=1.7$ K. The transition between the $TI$ turbulence to $TII$ turbulence   is clearly visible in the elbow of the plot.}\label{narrow4}
\end{figure}

For the critical Reynolds number $Rey_1$ considered,  the critical radius corresponding to the applied heat currents  $\dot Q=5\times 10^{-4}$ J/s and $\dot Q=10\times 10^{-4}$ J/s are respectively: $r_c=660\, \mu$m  and $r_c=1330\, \mu$m (for $T=1.5$ K); $r_c=750\, \mu$m  and $r_c=1500\, \mu$m (for $T=1.6$ K); $r_c=440\, \mu$m  and $r_c=880\, \mu$m (for $T=1.7$ K). In Figures \ref{narrow3} and \ref{narrow4}, the corresponding numerical results found in our model have been plotted for $d=50\;\mu$m. Heat transfer in He II in tubes  of this order of size has been explored experimentally in \cite{GBF_AIP}.

These estimations as well as the graphics plotted in Figure
\ref{narrow2} show that the higher is
the heat current  the higher
is the radius corresponding to the transition between all these regimes. Furthermore, the
effective thermal conductivity  drops down in the TI turbulent regime and much more
in the TII regime. Thus, in order to cool down the temperature of
a device, it is convenient to choose channels which show an
effective thermal conductivity closer to the Landau's estimations.
For a fixed heat current, this means that the radius of the channels
has to be large enough in order to be in $L=0$ regime or in the TI
regime. Another observation is that these features depend on the
temperature, and graphics show that lower temperature are
preferable.

\begin{figure}\label{narrow0}
  \includegraphics[width=10cm]{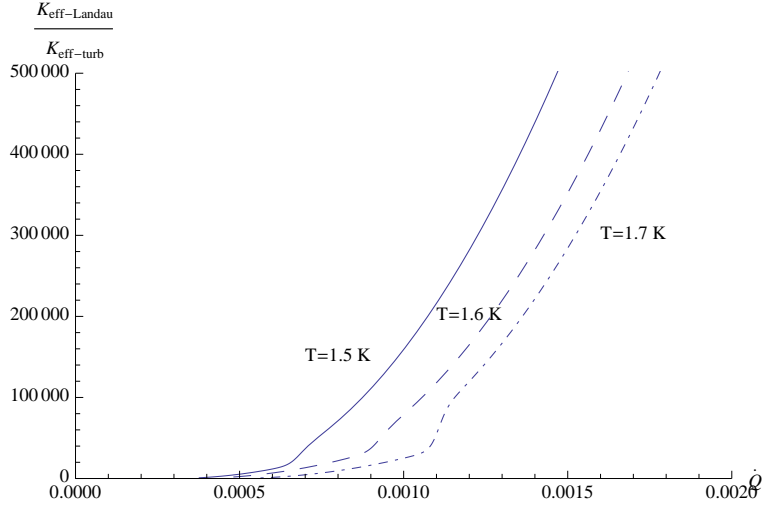}\\
  \caption{The behaviour of the ratio between thermal conductivity in superfluid
  helium without and with vortices  (\ref{conductivityNarr}) against the applied heat current $\dot Q$  (J/s) in the Martin and Tough's experiment for  diameter $d=1000\;\mu$m at three different temperatures,
   $T=1.5$ K (solid line), $T=1.6$ K (dashed line) and $T=1.7$ K (dot-dashed line). The transition between the $TI$ turbulence to $TII$ turbulence   is clearly visible in the elbow of the plot.}\label{narrow1}
\end{figure}

\section{Conclusions}
\setcounter{equation}{0}

In Section 3 of this paper we have established theoretically a   more complete relation between $\Delta T$ and $\dot Q,$
and the corresponding effective thermal conductivity
of microchannels filled with He II in
absence of net mass convection (counterflow situation). To do
that, we have taken into consideration the resistance force due to
quantized vortex tangle arising for high enough heat flux and, instead of assuming that $L\sim q^2$, as it follows from Vinen's equation\;(\ref{Vinen}), we have taken more detailed approach. This is based on equation (\ref{Vinen-gen2}), generalizing (\ref{Vinen}) through a term related to the influence of the walls on the vortex lines. This equation, and the behavior (\ref{L0}), (\ref{L1}) for their coefficients, yield an explicit mathematical model for the effective thermal conductivity of He II between Landau and Gorter-Mellink regimes. Unfortunately, this is not a simple expression, but it allows to bridge the gap between the two mentioned well-known regimes by means of an explicit physical model. 

This method cannot be extended to arbitrary narrow channel, because when the width of the channels becomes comparable or smaller than the mean free path of the heat carries, the regime becomes
ballistic and phonon collision with the walls, rather than with rotons or with themselves, become  the dominant factor \cite{Tabeling, Schmidt_ZPCM36}. Another aspect limiting the flow of heat from solid walls to liquid helium is the Kapitza conductance of the interface \cite{Snyder_Cry1970} but we do not deal with this topic here.

It is clearly shown in Figure\,\ref{narrow2}, that the effective thermal
conductivity is reduced more drastically in narrow channels  than in
wide channels. Thus, from a practical perspective, our paper indicates that, for a
given amount of helium, better contact is achieved between a solid
and the bulk He II if the contact is made through a few wider channels rather than many
narrow channels.  This is already known and experimentally confirmed \cite{Sciver2012, Arp_Cryo10}. For instance, for a certain given heat flux density, a large channel or ten small channels are used, all of them in laminar regime, the temperature difference is larger in the large channel (proportional to the square of the radius). However, this is no longer so in the turbulent regime. Thus, an understanding of the transition regime may be especially useful in cooling systems with high heat loads. 

This may be useful for practical purposes. In particular, the present analysis
shows the interest of equation  (\ref{Vinen-gen2})  generalizing the usual
Vinen's equation to take into account the effects of the walls. Such equation
describes in a natural way the transition from laminar to turbulent TI state  and, if it is complemented
with (\ref{gamma0}) and (\ref{omega}),  it also describes the transition from TI to TII turbulence. As it has been shown here, equation
 (\ref{Vinen-gen2}) is deeply related to the more practical topic of the effective thermal conductivity of narrow tubes filled with He II \cite{Sciver2012}.

\section*{Acknowledgements}
The authors acknowledge the support of the Universit\`{a}
di Palermo (under Grant Nos. Fondi 60\% 2012 and Progetto CoRI 2012, Azione d) and the collaboration agreement
between Universit\`{a} di Palermo and Universit\`{a}t Aut\`{o}noma
de Barcelona.
DJ acknowledges the financial support from the Direcci\'{o}n
General de Investigaci\'{o}n of the Spanish Ministry of Education
under grant FIS2009-13370-C02-01
 and of the Direcci\'{o} General de Recerca of
the Generalitat of Catalonia, under grant 2009 SGR-00164.
M.S. acknowledges the hospitality of the "Group of Fisica Estadistica of the Universit\`{a}t Aut\`{o}noma de Barcelona".

\end{document}